# Discrimination between spin-dependent charge transport and spin-dependent recombination in π-conjugated polymers by correlated current and electroluminescence-detected magnetic resonance

Marzieh Kavand[1], Douglas Baird[1], Kipp van Schooten[1•], Hans Malissa[1],

John M. Lupton[1,2], and Christoph Boehme[1]

[1] Department of Physics and Astronomy, University of Utah, Salt Lake City, Utah

[2] Institut für Experimentelle und Angewandte Physik, Universität Regensburg, D-93040 Regensburg, Germany

• Now at Dartmouth College, Hanover, New Hampshire, USA

**Abstract**

Spin-dependent processes play a crucial role in organic electronic devices. Spin coherence can give rise to spin mixing due to a number of processes such as hyperfine coupling, and leads to a range of magnetic field effects. However, it is not straightforward to differentiate between pure single-carrier spin-dependent transport processes which control the current and therefore the electroluminescence, and spin-dependent electron-hole recombination which determines the electroluminescence yield and in turn modulates the current. We therefore investigate the correlation between the dynamics of spin-dependent electric current and spin-dependent electroluminescence in two derivatives of the conjugated polymer poly(phenylene-vinylene) using simultaneously measured pulsed electrically detected (pEDMR) and optically detected (pODMR) magnetic resonance spectroscopy. This experimental approach requires careful analysis of the transient response functions under optical and electrical detection. At room temperature and under bipolar charge-carrier injection conditions, a correlation of the pEDMR and the pODMR signals is observed, consistent with the hypothesis that the recombination currents involve spin-dependent electronic transitions. This observation is inconsistent with the hypothesis that these signals are caused by spin-dependent charge carrier transport. These results therefore provide no evidence that supports earlier claims that spin-dependent transport plays a role for room temperature magnetoresistance effects. At low temperatures, however, the correlation between pEDMR and pODMR is weakened, demonstrating that more than one spin-dependent process influences the optoelectronic materials properties. This conclusion is consistent with prior studies of half-field resonances that were attributed to spin-dependent triplet exciton recombination which becomes significant at low temperatures when the triplet lifetime increases.

**PACS:**

| | |
|---|---|
| **71.20.Rv** | **Polymers and organic compounds** |
| **76.30.−v** | **Electron paramagnetic resonance and relaxation** |
| **73.61.−r** | **Electrical properties of specific thin films** |

## 1. Introduction

There has been considerable interest in understanding the exact physical nature of the spin-dependent processes responsible for the magneto-optoelectronic properties of organic semiconductor materials [1-4] such as $\pi$-conjugated polymers. Ultimately, these processes are crucial for the performance and efficiency of devices such as organic light-emitting diodes (OLEDs), solar cells, and potential sensor applications [5-8]. For typical room temperature device operating conditions, it has been established in recent years that weakly spin-exchange and spin-dipolar coupled pairs of charge carriers with spin s=1/2 are responsible on their own for spin-related material effects such as DC magnetoresistance [9-11] or conductivity changes due to magnetic resonant excitation by a radiofrequency field [12]. One of the crucial experiments in this context was the demonstration that spin-dependent electric currents under strong coherent magnetic resonant excitation of the charge carrier spins reveal characteristic beat oscillations due to Rabi-nutation of weakly coupled pairs of s=1/2 spin states [13-15]. For room-temperature conditions, this observation appears to be universal for poly(phenylene-vinylene) (PPV) based materials [13], fullerenes [16, 17], small molecule devices based on phthalocyanine [18], or blend materials [19,20]. This universal observation of spin beating conflicts with earlier hypotheses that processes involving higher spin-manifolds such as triplet excitons (s=1) or trions (s=3/2) could be responsible for the observed room-temperature spin-effects [21-24], since those spin manifolds would be expected to display Rabi-nutation frequencies that have not been observed in experiment.

In spite of this progress, controversy has remained regarding the polarity of the charges bound in the observed weakly spin-coupled charge-carrier pairs. Some studies have claimed that these carriers should be unipolar in the form of adjacently localized pairs of either electrons or holes, whose hopping or tunneling probability into so-called bipolaron states depends on their spin-pair permutation symmetry [25-31]. It is important to note that the term bipolaron here refers to the double occupation of electron states and not to the charge states, bipolarons are in fact electrostatically unipolar. Other studies have concluded that these pairs are always of bipolar nature in the true sense of bipolarity in that they are pairs of charge carriers with opposite charge, namely electrons and holes, forming so-called polaron pairs [32-39]. This ambiguity with regards to charge polarity has remained unresolved. Among the reasons for this controversy is that spin spectroscopy techniques are inherently insensitive to electrostatic effects, owing to the circumstance that the spin of a hole, which is an electron that is missing from a highest occupied molecular orbital state, is generally indistinguishable from the spin of an electron. The spin of a hole and thus the magnetic moment of a hole is therefore provided by the single electron that previously occupied the hole state. For any observable that is sensitive to paramagnetic states and their interactions, electron spins and hole spins are therefore of nearly identical physical nature and a theoretical treatment of the problem [40] has shown that, in principle, both the polaron-pair model as well as the bipolaron hypothesis can support experimentally observed magnetoresistance data. Similarly, coherent spin-motion experiments such as pulsed electrically or optically detected magnetic resonance (pEDMR or pODMR) are not able to resolve this problem directly for the same reason. As we will show below, both techniques have to be combined.

In recent years, attempts have been made to obtain insights into the polarity of the charges of weakly coupled spin pairs by consideration of both resonance spectroscopies. For instance, the study of beat oscillations of charge-carrier spin-Rabi nutation in the conductivity of diode structures based on poly[2-methoxy-5-(2-ethylhexyloxy)-1,4-phenylenevinylene] (MEH-PPV) has been repeated in optical recombination by pODMR spectroscopy through transient detection of photoluminescence intensity changes [41]. This comparison revealed qualitatively identical observations to pEDMR [14] of a pronounced spin-dependent transition rate which displays the same double Gaussian resonance lines, the same Rabi nutation of spin manifolds with s=1/2 as well as distinct beat oscillations, indicative that the observed spin systems belong to weakly coupled pairs with negligible dipolar and exchange coupling [42] [15]. While this experiment unambiguously proves that weakly coupled electron-hole pairs are involved in recombination, and thus luminescence, under the low-temperature conditions where measurements were made, the results do not unambiguously prove that the weakly coupled room-temperature charge carrier spin pairs observed with pulsed EDMR are also of opposite charge and not bipolarons.

The experiments presented in the following aim to resolve the question of whether the charge-carrier species in $\pi$-conjugated polymers that control spin-dependent electrical currents and those which control spin-dependent signatures in the luminescence are of the same or of differing physical nature. We have carried out experiments on several OLED devices where the transient current and electroluminescence response to a pulsed magnetic resonance excitation of charge carriers was measured by *simultaneous* detection of both observables – current and electroluminescence (EL). The experiments were conducted at several temperatures, as different temperatures influence the dynamics of both the electrically and optically detected signals due to the temperature dependance of pair recombination, dissociation and charge carrier mobility [35].

Measuring the correlation of electrically and optically detected spin-dependent signal *dynamics*, i.e. determining the Fourier components in the response to spin manipulation, allows us to discriminate between processes which affect both optical and electrical properties in the same way and those which only influence EL or conductivity. If the dynamic components of spin-dependent signals from both detection channels are highly correlated, it is very likely that both signals originate from the same underlying spin-dependent process. However, if the dynamics of electrically and optically detected spin signals show little correlation, then one may conclude that multiple spin-dependent processes are present in the material and that at least one of the processes affects the optical emission and electrical conductivity in different ways. Furthermore, spin-dependent transport processes will affect optical emission only indirectly by changing the device current and thus, as discussed quantitatively below, are separable from spin-dependent recombination. In this case, device current changes will, of course, also induce recombination rate changes. An example of a spin-dependent transport mechanism which has previously been resolved directly in MEH-PPV devices is the quenching interaction between triplet excitons and charge carriers, the so-called triplet exciton-polaron mechanism [14,43-49]. This process depends on triplet density and triplet lifetime, and therefore on temperature and exhibits clear signatures of s=1 triplet species in the half-field magnetic resonance. Data sets exhibiting only one spin-dependent process, with equal temporal dynamics under both optical and electrical detection, will make it possible to unambiguously confirm or refute whether magnetic resonantly manipulated electronic processes are spin-dependent transport or due to recombination mechanisms: changes in EL which cannot be accounted for by a change in sample current must always come from a recombination process. Particularly for the case of room-temperature experiments, where it is known that all spin-dependent processes observed by pEDMR are caused by weakly spin-coupled pairs of charge carriers [14], the experiment presented here is able to provide insight into the relative role of the polaron-pair mechanism compared to the bipolaron mechanism.

The idea behind this study follows from previous pEDMR and pODMR correlation studies of spin-dependent processes in hydrogenated amorphous silicon [50] and silicon nitride [51]. In contrast to these studies, the pEDMR and pODMR experiments presented here, on two types of PPV, were recorded simultaneously on the same device. This approach is a prerequisite for the correlation study because of the substantial random variations between materials and between individual OLED devices. Furthermore, given that EL but not photoluminescence is detected simultaneously with the current, in the absence of photo-induced charge carriers, the presence of geminate and thus spin-correlated recombination can be excluded.

### 2. Comparing the dynamics of pEDMR and pODMR signals

In order to study the correlation between the dynamics of magnetic resonance induced current and EL changes, we consider the interdependency of the photon current $J_\gamma$ (i.e. the EL emission rate) and the electrical injection current $J_E$ of the device. The latter can be adjusted by changing the device bias and thus by choice of the overall device operating point. The ratio between the photons emitted from the device (which is a function of the injection current) and the injection current itself, $J_\gamma(J_E)/J_E$ is the static quantum efficiency, which, like both $J_\gamma$ and $J_E$, is a function that varies with the device operating point. For most OLED devices, this variation is only weak, mostly towards higher biases. The parameter $\alpha = \partial J_\gamma / \partial J_E$ is the dynamic quantum efficiency, which represents the small change in EL intensity when a small change in device current occurs. In particular, a small change $\delta J_E$ to the device current, e.g. due to a small change in device bias, will lead to a change of the EL intensity

$$\delta J_\gamma = \frac{\partial J_\gamma}{\partial J_E} \delta J_E = \alpha \delta J_E .$$ 

(1)

At small forward bias, when the OLED has not turned on, $\alpha = 0$. As the device current approaches its turn-on operating point, $\alpha$ becomes strongly bias-dependent and thus a device current dependent function. Subsequently, for biases beyond this regime, $\alpha$ usually becomes constant again, independent of bias and device current, yet it is non-zero, indicating a linear relationship between the emitted EL intensity and the applied device current. The slope $\alpha$ in this regime is then determined by the radiative and non-radiative charge-carrier recombination probabilities, the electron-hole pair generation rates, and various other electronic transition rate coefficients which determine the electronic processes in the device studied [35]. Figure 4(d) which is discussed below in detail shows an example of a room temperature function $J_\gamma(J_E)/J_E$ that was measured on a SY-PPV based OLED device. It shows good agreement with a linear function, implying that $\alpha = const.$, over almost the entire displayed range.

Following Eq. 1, we realize that a change $\delta J_E$ of the device current due to a magnetic resonance induced change of spin-dependent charge transport will leave $\alpha$ unchanged ($\delta\alpha = 0$), yet they will cause a change of radiative recombination $\delta J_\gamma$, such that

$$\frac{\delta J_\gamma / \delta J_E}{\partial J_\gamma / \partial J_E} = \frac{\delta J_\gamma / \delta J_E}{\alpha} = 1 \, . \tag{2}$$

The independence of $\alpha$ from the operating point is maintained as long as heating and device degradation are not significant, and also, as long as magnetic resonantly induced changes of spin-dependent recombination rates do not modify the ratio between radiative (singlet) and non-radiative (triplet) recombination. If the latter occurs, as in EDMR or ODMR experiments, $\alpha$ is changed, typically by a small amount $\delta\alpha$, with a magnitude $|\delta\alpha| << \alpha$, causing a magnetic resonantly induced change of the EL emission $\delta J_\gamma$ (the ODMR signal) as well as a change of the device current

$$\delta J_E \approx \frac{\delta J_\gamma}{\alpha} = \frac{\delta\alpha}{\alpha} J_E \, , \tag{3}$$

which is the EDMR signal due to a change of the overall recombination current. Note that Eq. 3 neglects very small second order contributions of current changes caused by recombination changes induced by the current changes themselves as these will be on the order of $(\delta\alpha/\alpha)^2$. Since in the case of modified recombination probabilities the EL emission rate $J_\gamma$ is influenced by both changes $\delta\alpha$ of the dynamic quantum efficiency $\alpha$ as well as changes $\delta J_E$ of the injection current $J_E$, Eq. 1 assumes the form

$$\delta J_\gamma \approx \alpha \delta J_E + \delta\alpha J_E \tag{4}$$

for which, again, a negligibly small second-order contribution $\delta\alpha\delta J_E$ is removed and therefore, utilizing Eq. (3) and (4),

$$\frac{\delta J_\gamma / \delta J_E}{\partial J_\gamma / \partial J_E} = \frac{\delta J_\gamma / \delta J_E}{\alpha} \approx 1 + \frac{\delta\alpha}{\alpha} \frac{J_E}{\delta J_E} \approx 2 \tag{5} .$$

In summary, when a spin-dependent *transport rate*, which is changed by magnetic resonance, causes a current change $\delta J_E$, the resulting change of the EL $\delta J_\gamma$ is solely due to the changed electrical current but not due to a changed recombination probability. Thus $\delta\alpha = 0$, while $\left(\delta J_\gamma / \delta J_E\right)/\left(\partial J_\gamma / \partial J_E\right) = 1$, according to Eq. 2. In contrast, if the *recombination rate* changes due to the magnetic resonance excitation, Eq. 5 applies and $\left(\delta J_\gamma / \delta J_E\right)/\left(\partial J_\gamma / \partial J_E\right) \approx 2$. The deviation of Eq. 5 from the value of 2 will always depend on how significant second order contributions will be, which were neglected in the derivation of Eq. 5. In general though, $\left(\delta J_\gamma / \delta J_E\right)/\left(\partial J_\gamma / \partial J_E\right) \neq 1$ implies that $\delta\alpha \neq 0$ and thus that spin-dependent recombination is involved in an observed pEDMR and pODMR signal.

The term $\left(\delta J_\gamma / \delta J_E\right)/\left(\partial J_\gamma / \partial J_E\right) = \left(\delta J_\gamma / \delta J_E\right)/\alpha$ in Eqs. 2 and 5 represents the ratio of the measured absolute changes of the EL emission rate and the current, normalized by the dynamic quantum efficiency $\alpha$. While it is experimentally challenging to measure absolute photon emission rates, it is straight forward to measure this ratio because this normalization is neither affected by the finite collection efficiencies of any realistic optical detection setup nor by spin-independent charge currents which typically shunt spin-dependent electric currents in realistic devices. Thus, Eqs. 2 and 5 still hold even if $\alpha$ does *not* represent the actual dynamic quantum efficiency but only the derivative of the *detected* photon flux over the *applied* electric current in the presence of large photon losses and shunt currents.

The arguments made for Eqs. 2 and 5 apply equally for the dynamic case when harmonic oscillations of $J_\gamma$ and $J_E$ with frequency $\omega$ are induced by magnetic resonance. As long as $\delta\alpha$, and $\alpha$ depend on $\omega$ in the same way, i. e. Eq. 3 holds, independently of the frequency $\omega$, it can even be applied to the comparison of transient ODMR and EDMR signals by comparing the magnitudes $\left|FT\{\delta J_\gamma\}(\omega)\right|$ and $\left|FT\{\delta J_E\}(\omega)\right|$ of the Fourier components of the recorded transients $\delta J_\gamma(t)$ and $\delta J_E(t)$ such that

$$\frac{\left|FT\{\delta J_\gamma\}(\omega)\right|/\left|FT\{\delta J_E\}(\omega)\right|}{\partial J_\gamma / \partial J_E} := \xi(\omega) \quad \begin{cases} = 1 & \text{for spin-dependent transport} \\ \neq 1 & \text{for spin-dependent recombination} \end{cases} \quad (6).$$

Following Eq. 6, we realize that:

(i)     When simultaneous pEDMR and pODMR experiments reveal $\xi(\omega) = C$, with $C \neq 0$ being a constant value that is independent of frequency $\omega$, then the underlying dynamics of the electrically and optically detected spin signals follow identical dynamical behavior and are likely due to the same microscopic electronic processes. While this case allows the discrimination between the situation where conductivity and EL are governed by the same processes versus the two observables being controlled by several different processes, this discrimination alone cannot allow us to distinguish absolutely between spin-dependent recombination and spin-dependent transport.

(ii)    When the pEDMR and pODMR experiments show that $\xi(\omega) \neq 1$, then $\delta\alpha(\omega) \neq 0$ which implies that the observed changes in the EL *deviate* from the changes anticipated solely from the resonantly induced electric current changes. Therefore, spin-dependent transport effects cannot account for these observations on their own. Consequently, when $\delta\alpha(\omega) \neq 0$ spin-dependent recombination must contribute to the observed spin-dependent electrical and optical processes. If spin-dependent transport arises in addition and modifies the current and thereby the EL intensity, it is highly unlikely to follow exactly the same temporal dynamics. In this case $\xi(\omega)$ will not be independent of $\omega$.

(iii)   (i) and (ii) also imply that if $\xi(\omega) = C > 1$ and independent of $\omega$ then spin-dependent recombination is likely to be the sole contributing process to pEDMR and pODMR.

We note that while the absolute magnitude of $\alpha$ is not relevant because of the normalization, the applicability of Eq. 6 requires that $\alpha$ be independent of $\omega$. Since $\alpha$ depends on the charge and photon detection sensitivities, which depend on the bandwidth of the instrumentation used for detection, $\alpha$ can be constant only within a limited region of frequencies. Finding this region for the two detection pathways is accomplished by careful characterization of the dynamics of both the current and photon detection and is a crucial prerequisite for the interpretability of simultaneous pEDMR/pODMR experiments. Finally, we note that while $\delta\alpha(\omega) \neq 0$ implies the involvement of recombination in the observed spin-dependent signals, it does not imply $\delta\alpha(\omega) > 0$, i.e. that under magnetic resonant excitation, the optically detected radiative recombination rate increases. In fact, $\delta\alpha(\omega)$ may be positive or negative, even when the net charge carrier recombination rate increases. This ambivalence arises since the sign of both the optically detected radiative recombination rate change as well as the non-radiative recombination rate change – which is mostly due to recombination of triplet excitons and thus remains undetected – depend on various parameters which also control the changes in spin-dependent electronic transitions. For the example of charge-carrier pair processes, these parameters include the singlet and triplet recombination probabilities of charge carrier pairs, the pair dissociation probabilities, the longitudinal and transverse spin-relaxation times of electrons and holes, the distribution of hyperfine fields, and the spin statistics of the charge-carrier pair generation rates. These dependencies have been discussed in detail elsewhere [35, 50] and they imply that the signs of the observed pEDMR or pODMR signals do not allow for any unambiguous conclusion about the qualitative or quantitative nature of the underlying spin-dependent process.

### 3. Experiments

#### (i) Setup

We carried out pulsed magnetic resonance spectroscopy on OLEDs as described in detail in Ref. [14,50]. In Figure 1(a) the sketch of the experimental setup for carrying out simultaneous measurements of pEDMR and pODMR on OLED structures is shown. A sample holder was designed specifically for these experiments to allow for both electrical connection to the sample and optical collection of the EL emission. The sample holder has a printed circuit board (PCB) at its tip for electrical contact. A prism with 2 mm length was placed on top of the PCB to couple the EL into a bundle of four optical fibers. Figure 1(b) illustrates the prism with collection fibers and Fig. 1(c) shows a photograph of the tip of the sample holder with an operating OLED in place. The OLED structure is given in panel (d). The sample holder is designed in such a way that the sample is automatically positioned in the center of a cylindrical dielectric microwave resonator. A constant voltage bias was applied to the device and the electrical current and EL were recorded simultaneously. To use magnetic resonance to manipulate the spins which influence conductivity and EL, a magnetic field was applied to induce a Zeeman splitting of the spin states. A short X-band microwave pulse with a duration of 400 ns was applied and the transient variations in current and EL were recorded simultaneously. For pEDMR detection, a Stanford Research SR570 current amplifier was used for current-to-voltage conversion, amplification and filtering of the signal. For pODMR experiments, where EL emission is detected, light collected by the fibers was directed through a collimator lens where it was subsequently focused onto a Femto LCA-S-400K-SI photodiode. The output of the photodiode was amplified and filtered by a Stanford Research SR560 voltage amplifier. Both the voltage and the current amplifiers had specific frequency settings which were determined by band-pass filters as discussed below. Two separate yet identical analog to digital converters (ADCs) were then used to digitize the pEDMR and pODMR transients.

(ii)     Current and EL transients

Figure 2 shows simultaneously recorded pEDMR and pODMR transients of an OLED employing the commercially available so-called "super-yellow" PPV derivative SY-PPV. The transients are plotted in panels (a) and (b) on color scales showing relative changes in current ($\delta J_E/J_E$) and EL intensity ($\delta J_\gamma/J_\gamma$), respectively, as functions of the magnetic field strength. The data sets display clear resonance features around a magnetic field of 343 mT as indicated by the dashed lines.

Figure 2(c) compares both transients directly at the field of maximum resonance at $B_0 = 343$ mT, while panel (d) compares the magnetic field dependencies of the normalized current change $j_E^{norm}(B_0)$ and normalized EL change $j_\gamma^{norm}(B_0)$ recorded at times $t_E = 25.24$ μs and $t_\gamma = 8.87$ μs, respectively, when their signal maxima arise. Both are indicated by vertical dashed lines in panels (a) and (b).

As expected from previous room temperature pEDMR experiments, conducted in absence of photo excitation [52], the resonant excitation causes a quenching of the device current. While this observation is consistent with a recombination rate increase causing a conductivity decrease, as explained above these results constitute no proof whatsoever for this since both spin-dependent recombination *and* transport rates can cause either a current enhancement *or* decrease. Also, as expected, given that $\alpha > 0$, the decrease of the current shown in (a) causes a decrease of the optical output in (b). However, in this particular example, the relative change of the optical output is significantly larger. This observation alone is not proof for or against the optical output being governed solely by the current change and not by changes of the recombination rate.

Figure 2(d) confirms what previous measurements of pEDMR and photoluminescence detected pODMR have shown [13, 41]: at room temperature, the resonances of both electrical and optical detection channels reveal identical spectral features within the given noise level. The similarity between the two spectra can be seen clearly from the difference of the two signals displayed in the inset. As discussed previously [13, 53, 54], the pODMR as well as the pEDMR spectra are well represented by two Gaussian resonance lines (solid black lines). While from this data, together with previous studies [13, 41, 54], we know that the device conductivity and the EL intensity are controlled by the same two spin manifolds – those of weakly spin-coupled electrons and holes – the observations are, again, neither a proof that these are the same spin-dependent processes, nor that these signals are due to either one particular or several spin-dependent processes at the same time. These resonances also do not prove or disprove that the observed EL transient is solely a consequence of the dynamic change in conductivity. This ambiguity is why only a complete dynamical analysis of these spin-dependent signals can give further insight into the underlying nature of these observations, as presented in the following.

The comparative plot of the dynamics of the two signals shown in Fig. 2(c) suggests that the pEDMR transient exhibits a different, much slower dynamics than the pODMR transient. One could therefore be tempted to conclude that pEDMR and pODMR originate from separate spin-dependent signals with different dynamical behavior. However, since we are comparing two fundamentally different observables it is crucial to take into account the way these signals are detected. The SR570 current amplifier was used for pEDMR detection while a photodiode and a SR560 voltage amplifier were employed for the pODMR measurement. If the two different detection schemes do not exhibit identical dynamical behavior, the detection system with the higher bandwidth will display a faster transient, even if the signals detected by each separate channel have identical dynamical characteristics.

Following the discussion of Eq. 6 above, the data sets in Fig. 2 actually demonstrate how different detection systems with different dynamic behavior cause a frequency dependence of the dynamic quantum yield $\alpha$ and thus show how the direct, uncorrected comparison of the two raw time transients with respect to Eq. 5 is misleading. Instead, the measured data from each detection channel must be corrected in a way that effectively *removes* the influence of the frequency response of the respective detection system before a comparison of the pODMR and pEDMR dynamics is possible. For this, the frequency response (i.e. the transfer function) of the individual electrical and optical detection pathways must first be established.

(iii)    Determination of detector transfer functions

In order to assess the effects of the equipment and the respective filter settings on the detected signals and to outline the correction procedure needed, a series of pEDMR measurements was carried out with different settings of the frequency filter. Figure 3(a) shows the relative change of the electrical current ($\delta J_E/J_E$) as a function of time for three different current amplifier filter settings. All three measurements were carried out on the same SY-PPV device and under the same operating conditions (room temperature, a bias of 3.73 V, $J_E = 126.24(32)$ µA). To ensure reproducibility, the first measurement was repeated after finishing the last measurement. This confirmed that the same signal could be reproduced and the sample had not changed during the measurements. It is apparent that narrowing the bandwidth of the current amplifier distorts the transient. To account for this distortion, we consider the frequency components of the measured response. In Fig. 3(b) the magnitude of the Fast Fourier Transform (FFT) of the data in Fig. 3(a) is plotted. The frequency response of the different measurements clearly varies and, as expected, the broader bandwidth setting leads to the detection of higher-frequency signals. Thus, in order to compensate for the influence of the detection setup on the recorded signal, both the real and imaginary part of the transfer function of the detection setup is measured and used to process the raw data in the frequency domain.

To measure the transfer function related to each current amplifier setting, an alternating voltage of constant amplitude was applied to a resistor on the input channel of the current amplifier. The frequency of the applied AC voltage was then swept from 100 Hz to 10 MHz, while the differential voltage across the input resistor as well as the output response of the current amplifier were monitored in terms of magnitude as well as phase shift. Using these output and input measurements, the transfer function was calculated. Figure 3(c) plots the magnitude of the transfer function for the three different amplifier frequency-filter settings associated with the data in panel (a). These transfer functions were used to correct the FFT signals by deconvolution of the FFT data in panel (b) with the corresponding transfer function whose magnitudes are plotted in panel (c). The results of this procedure are plotted in panel (d) and show that all deconvoluted frequency distributions of the measured current transients are equivalent within the given noise levels. As expected, of course, a more selective bandwidth of the detection setup will lead to measurements with a higher noise level for frequencies outside of the respective amplifier bandwidth settings.

We applied a similar transfer function correction to all measured pODMR data. To measure the transfer function for the optical detection setup, the alternating voltage input signal in the pEDMR calibration was used to modulate a red LED coupled to the photodiode, which was connected to the voltage preamplifier. The output of the voltage preamplifier was then recorded as a function of the applied frequency. This approach ensured that the transfer dynamics of the entire setup were correctly measured, including the detector, preamplifier and main amplifier.

### 4. Results

(i)  Room temperature

Figure 4(a) displays the relative changes of the electrical current ($\delta J_E/J_E$) and the EL intensity ($\delta I_\gamma/J_\gamma$) as a function of time. These transients were measured simultaneously on the same SY-PPV OLED device after application of a 400 ns microwave pulse under the resonance condition (for g~2 at $B_0$ = 343.135 mT with a carrier frequency of the microwave pulse of 9.623216 GHz). The bandwidths for each of the two detection channels were chosen to be as large as possible, but at the same time as narrow as necessary so that the data could be recorded with an acceptable signal to noise ratio.

Figure 4(b) displays the magnitude of the *corrected* FFT of the relative changes in current $\left|FFT\{\delta J_E\}(\omega)\right|/J_E$ (blue curve) and EL emission $\left|FFT\{\delta J_\gamma\}(\omega)\right|/J_\gamma$ (red curve) of the data shown in panel (a), using the measured transfer functions plotted in Fig. 4(c). Given this data, we then recorded the detected EL intensity $J_\gamma(J_E)$ as a function of electric current $J_E$ as plotted in Fig. 4(d). A near ideal linear behavior close to the device operating point (indicated by the solid black line representing a linear fit) reveals the value of $\partial J_\gamma / \partial J_E$ and enables the comparison of the simultaneous pODMR and pEDMR measurements by taking the ratio $\xi(\omega)$ $=\left|FT\{\delta J_\gamma\}(\omega)\right|/\left|FT\{\delta J_E\}(\omega)\right|/\left(\partial J_\gamma / \partial J_E\right)$ into account as discussed above. Figure 4(e) displays a plot of this ratio as a function of the frequency $f = \omega / 2\pi$.

Despite having two entirely independent detection channels whose data sets were corrected by two entirely independent transfer functions, we find that $\xi(\omega)$ is constant within the given noise levels and independent of $\omega$, up to the frequency range where noise becomes too strong to make a meaningful determination of this ratio. Crucially, the ratio is not only independent of $\omega$, it is also found to be significantly larger than unity with $\xi(\omega) = C = 1.67(5)$ as shown by the horizontal solid black line, indicating a strong correlation between the corrected Fast Fourier components of the pODMR and the pEDMR signals.

In order to scrutinize this apparent correlation, correlation diagrams plotting the magnitudes of the Fast Fourier components $\left|FFT\{\delta J_\gamma\}(\omega)\right|$ of the pODMR signal with frequency $\omega$ against the magnitudes of the Fast Fourier components $\left|FFT\{\delta J_E\}(\omega)\right|$ of the pEDMR signals at the same frequency are shown in panels (f) to (i). The data in all four of these panels are based on the same Fast Fourier decomposition shown in panel (e), yet the four plots display this correlation up to different upper cut-off frequencies $\omega_c = 2\pi f_c$. Without any cut-off [panel (f)], two features are seen around which the data points scatter: a linear function that is close to the diagonal of the plot as well as a vertical feature. The vertical feature can be attributed to the different bandwidths of the two different detection channels ($f_c = 30$ kHz for the optical and $f_c = 500$ kHz for the electrical channel), implying that for Fast Fourier components of higher frequencies, there is much noise for the pODMR signal but little noise for the pEDMR signal. This interpretation is supported by panels (g), (h) and (i) which show the same data for decreasing upper cut-off frequencies and reveal that the vertical scatter is reduced when the cut-off frequency is decreased. The scatter disappears entirely when the cut-off frequency becomes similar to the low-pass frequency of the optical detection channel [see panel (i)]. For the following discussion of the pODMR to pEDMR correlation, we therefore only consider correlation plots $f_c = 50$ kHz. For room temperature SY-PPV, this is presented by Fig. 4(i), which reveals a high correlation between the dynamic behavior of pEDMR and pODMR signals, with a Pearson correlation coefficient of 0.94.

### (ii)    Temperature dependence

We repeated the simultaneous pEDMR/pODMR experiments at different temperatures and for for SY-PPV OLED devices as well as for MEH-PPV devices. Figure 5 displays the results of the SY-PPV experiments for device temperatures of 100 K, 20 K and 5 K, while Fig. 6 displays results for MEH-PPV experiments conducted at room temperature (290 K), 80 K, and 4 K. The measurements reveal that for MEH-PPV at room temperature, $\xi(\omega) = C = 2.92(5)$ is independent of $\omega$, as it is for SY-PPV at room temperature. In addition, this constant ratio is significantly larger than unity. For both polymer derivatives the ratio $\xi(\omega)$ is not frequency independent anymore for lower temperatures, indicating that the correlation between the dynamics of pEDMR and pODMR signal is reduced.

For all data sets in Fig. 5 and 6, we also display the corrected FFT of the transient EDMR and ODMR signals, $\xi(\omega)$, and the corresponding correlation plots. For SY-PPV, Pearson correlation coefficients of 0.94, 0.96, 0.84, and 0.59 are obtained for room temperature (290K), 100K, 20K, and 5K, respectively, while for MEH-PPV, 0.84, 0.99, and 0.87 are found for 290K, 80K, and 4K, respectively.

### 5.  **Discussion**

The measurements presented in Figs. 4 to 6 show that the changes of EL and conductivity of MEH-PPV and SY-PPV induced under magnetic resonance at room temperature and around 80-100K, exhibit a high level of correlation, indicating identical dynamic characteristics of the simultaneously recorded pEDMR and pODMR signals. In each of these materials, spin-dependent EL and conductivity are therefore likely governed by one single electronic process which affects both observables in the same way. Changes under resonance of the EL deviate significantly from what would be expected if they were caused solely by the resonantly changed device currents. Therefore, the observed changes in EL are due to both the modified device currents *as well as* changes of the probability of radiative charge carrier recombination of the pair. It is therefore concluded that in both materials studied, spin-dependent recombination is the cause of the observed electrically and optically detected spin-dependent signals. Since it has been well established that room temperature pEDMR and pODMR signals are due to pairs of weakly coupled spin manifolds with s=1/2 [6, 13, 53], it is therefore also concluded that the polaron-pair recombination mechanism is responsible for the observed signals.

While the experiments presented here do not conclusively allow us to rule out the influence of a spin-dependent transport process such as the bipolaron mechanism, they also provide no indication for its existence. The hypothesis that a significant spin-dependent transport channel exists implies that pEDMR signals caused by this process must exhibit a dynamical behavior identical to the observed spin-dependent recombination process as well as an identical magnetic resonance line shape, at room temperature. This would be highly unlikely given the conditions necessary (i.e. equal dissociation and recombination kinetics, coherence and relaxation times, and hyperfine field distributions for the carriers involved). We therefore conclude that there is unlikely any influence of spin-dependent transport on the observed electric current in the devices investigated here.

For SY-PPV, at low temperatures, the observed pEDMR and pODMR experiments display a modified and mutually different dynamic behavior where $\xi(\omega)$ is strongly dependent on $\omega$ for both studied materials. One can therefore conclude that EL emission and conductivity are either governed by entirely different electronic processes or by multiple processes, among which some may affect both detection channels while others may influence only one. While the measurements presented here do not allow further conclusions about the microscopic nature of these additional spin-dependent low-temperature processes, the observed behavior can be explained by the previously reported interaction between triplet excitons and lone charge carriers [14, 55] which affects the mobility of free charges without changing the radiative recombination rates. In this process, a spin-1 triplet exciton is annihilated by a spin-1/2 polaron in a spin-dependent process which shows resonance behavior of both spin multiplicities. For MEH-PPV, the change between room temperature behavior and low-temperature behavior appears to be analogous to SY-PPV. However, given the slight mismatch of the frequency range that is detectable with the used experimental setup (~0.5 kHz – 30 kHz) with the dynamics of the observed spin-dependent electronic transitions at room temperature, the available FFT data of MEH-PPV turned out to be noisier compared to the FFT data of SY-PPV. The resulting Pearson correlation coefficient is consequently smaller and thus, a monotonous decline of the correlation coefficient by decreasing temperature cannot be confirmed unambiguously.

## 6.  Conclusions

Simultaneously recorded pEDMR and pODMR experiments that detect electric currents and EL of OLEDs based on SY-PPV and MEH-PPV show that a single spin-dependent recombination process governs both observables at room temperature while no indication for the presence of spin-dependent transport is seen. Owing to the previously established fact that the observed recombination process is caused by weakly coupled spin pairs with s=1/2, it is concluded that the polaron-pair process is seen in these experiments. In contrast, at low temperatures, several spin-dependent processes govern the conductivity, consistent with the expected appearance of the triplet-exciton polaron quenching process. Again, no indications for the existence of the spin-dependent bipolaron transport process are found. These observations suggest that the majority of magnetoresistive and, more generally, magnetooptoelectronic effects in $\pi$-conjugated materials may be described by bipolar carrier-pair interactions.

## Acknowledgements

This work was supported by the US Department of Energy, Office of Basic Energy Sciences, Division of Materials Sciences and Engineering under Award #DE-SC0000909.

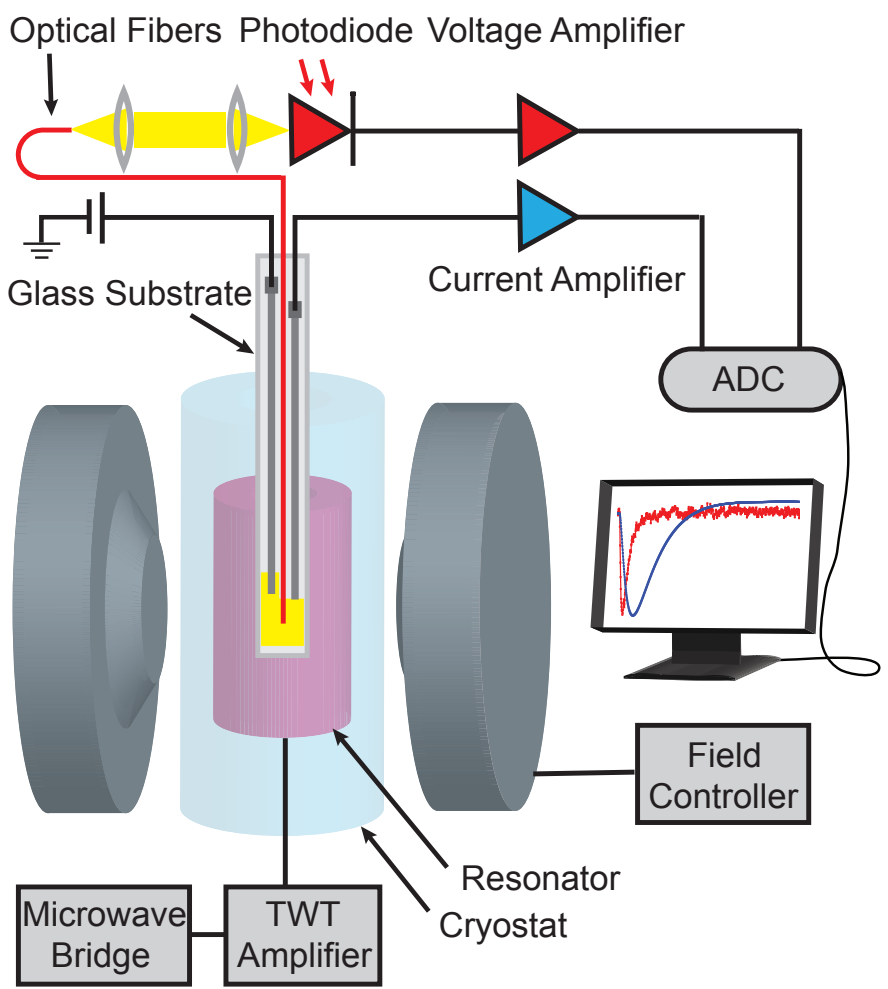

a) Optical Fibers — Photodiode — Voltage Amplifier

Glass Substrate

Current Amplifier

ADC

Microwave Bridge — TWT Amplifier — Resonator Cryostat

Field Controller

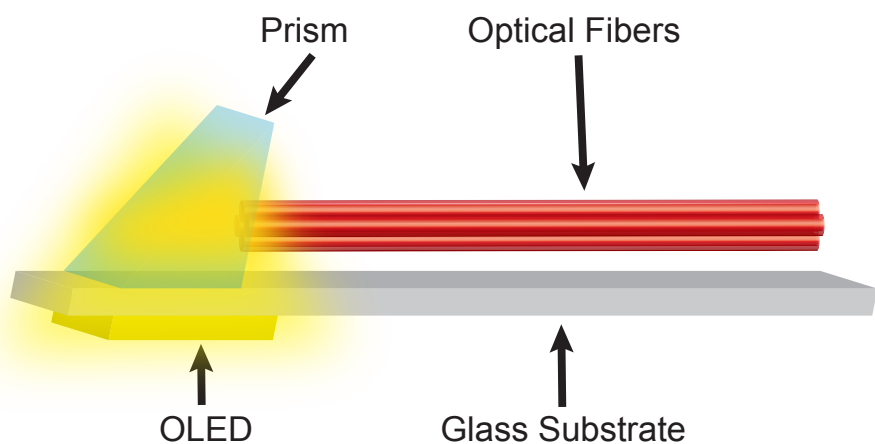

b) Prism — Optical Fibers

OLED — Glass Substrate

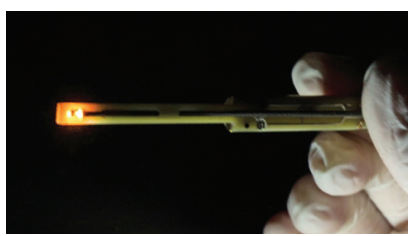

c)

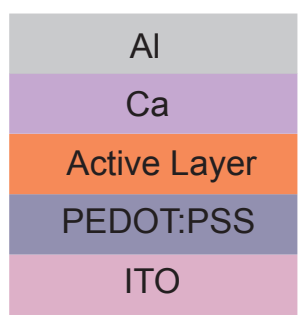

d)

| Al |
| Ca |
| Active Layer |
| PEDOT:PSS |
| ITO |

FIG. 1. (a) Schematic of the simultaneous pEDMR/pODMR experimental setup. (b) Sketch of the tip of the OLED sample holder. Both the current change and the electroluminescence (EL) change under resonance are recorded. A prism is used to couple the EL from the device into the optical fibers. (c) Photo of the sample under operation. (d) Structure of the OLED device, which is mounted on a narrow glass substrate with thin Al and ITO layers as electrical leads.

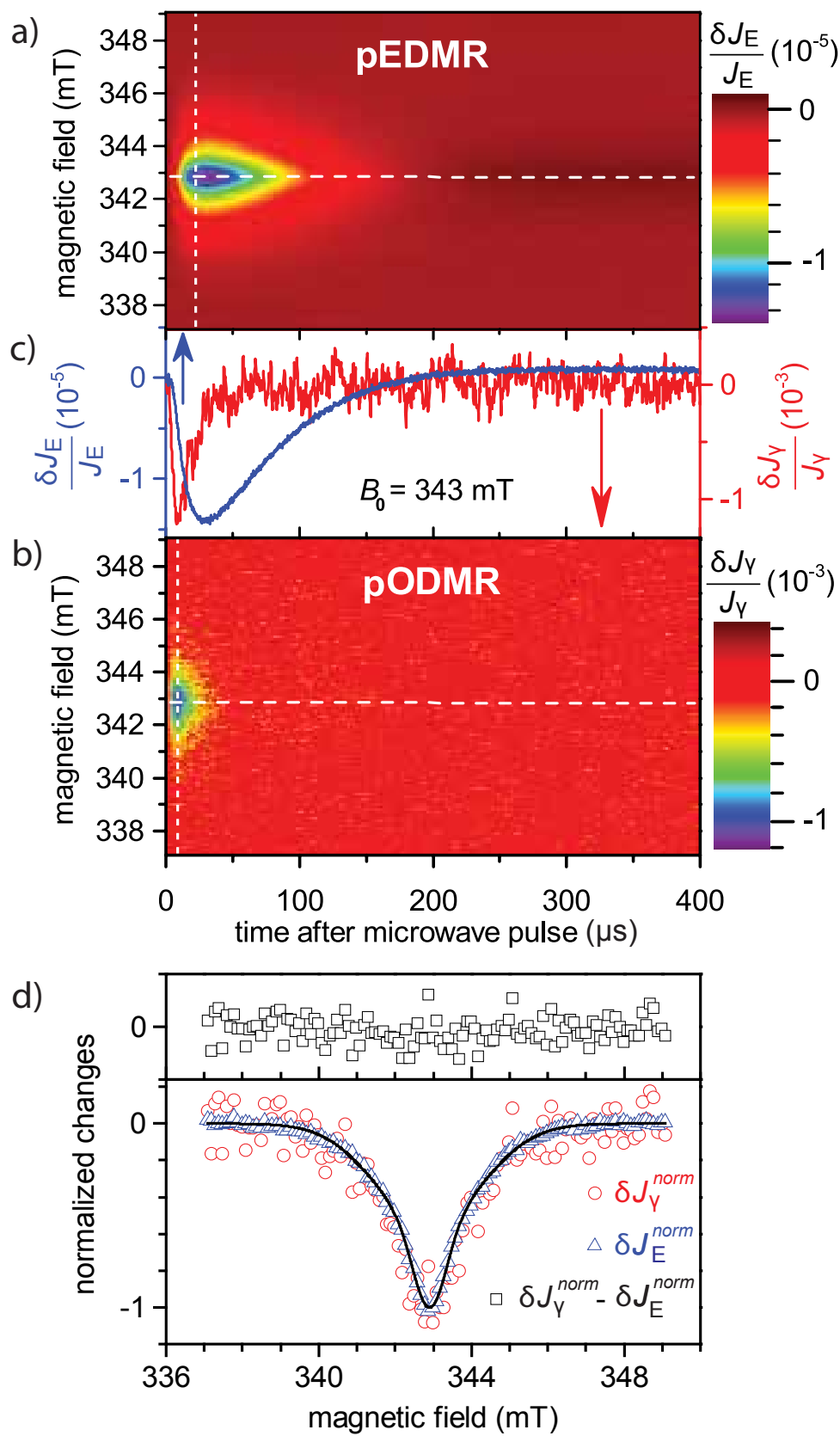

a) **pEDMR**

$\frac{\delta J_E}{J_E}$ ($10^{-5}$)

magnetic field (mT)

c)

$\frac{\delta J_E}{J_E}$ ($10^{-5}$)

$\frac{\delta J_Y}{J_Y}$ ($10^{-3}$)

$B_0$ = 343 mT

b) **pODMR**

$\frac{\delta J_Y}{J_Y}$ ($10^{-3}$)

magnetic field (mT)

time after microwave pulse (μs)

d) normalized changes

magnetic field (mT)

$\delta J_Y^{norm}$
$\delta J_E^{norm}$
$\delta J_Y^{norm} - \delta J_E^{norm}$

FIG. 2. Simultaneous measurement of pEDMR and pODMR in a SY-PPV OLED. A constant forward bias of 3.1 V was applied to the device. (a) Plot of the relative current changes $\delta J_E / J_E$ after a short microwave pulse of duration 400 ns and frequency 9.615308 GHz with microwave field amplitude $B_1 \approx 0.3$ mT was applied, as a function of time after the pulse and magnetic field strength $B_0$. (b) Plot of the differential change in EL intensity $\delta J_\gamma / J_\gamma$ measured simultaneously with the current change in panel (a). (c) Plot of the differential current (blue) and EL (red) emission transients under resonance at $B_0$ = 343 mT. The dynamics of the measured transients appear to differ. This is because the two different detection schemes for measuring changes in current and changes in EL, respectively, do not exhibit identical dynamical behavior. (d) Comparison of the current change normalized to the maximal current change $\delta J_E^{norm}(B_0) = \delta J_E(B_0) / \delta J_E^{max}$ taken from the data in (a) at the time $t_E$ = 25.24 μs (as indicated by a vertical dashed line) with the normalized differential change in EL $\delta J_\gamma^{norm}(B_0) = \delta J_\gamma(B_0) / \delta J_\gamma^{max}$ taken from the data in (b) at the time $t_\gamma$ = 8.87 μs (also indicated by a vertical dashed line) as a function of $B_0$. Both magnetic resonance line shapes display identical resonance distributions. The difference of the normalized spectra displayed in the upper subpanel in (d) reveals no significant residue. The black solid line represents a fit of $\delta J_E^{norm}(B_0)$ with a double Gaussian function representative of the hyperfine field distributions of the two carriers in the pair.

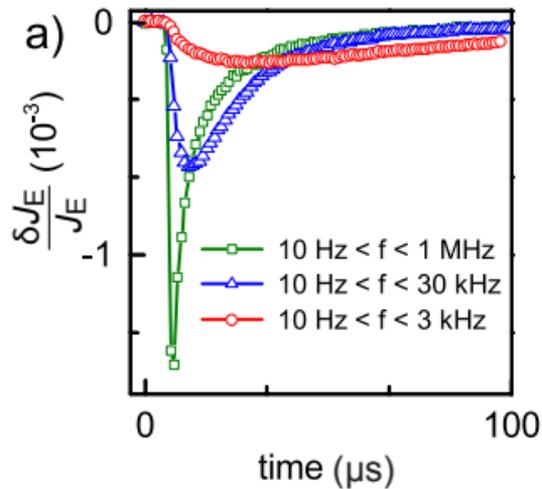

a)

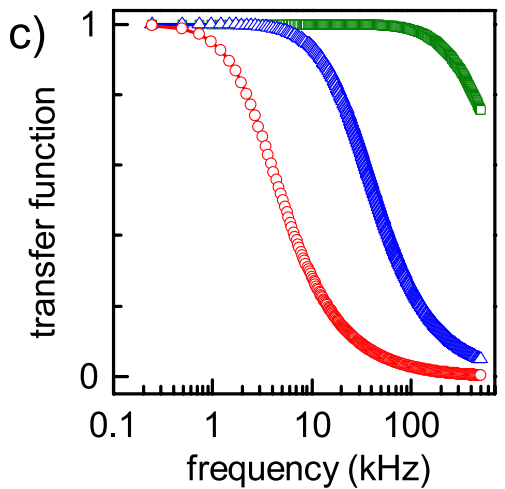

c)

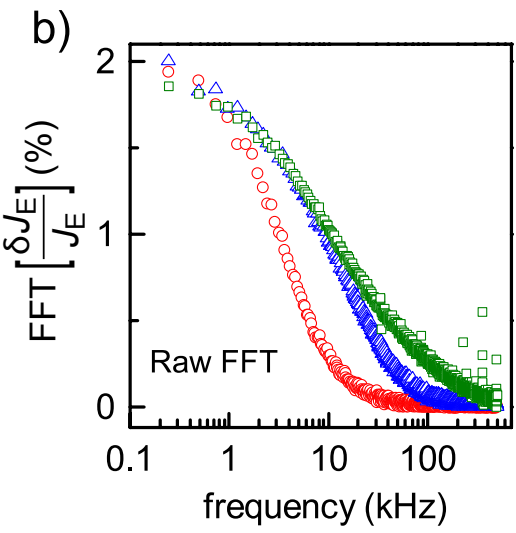

b)

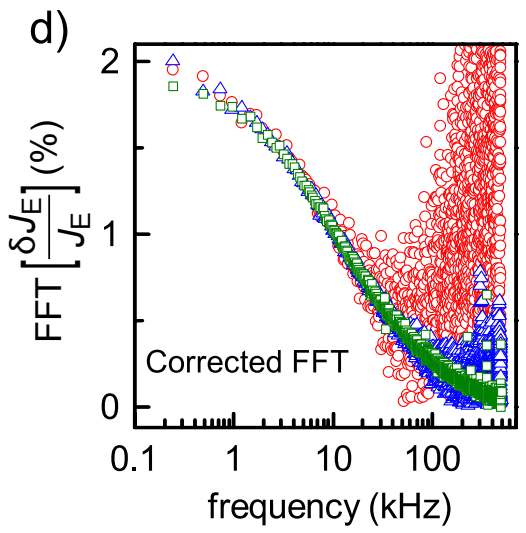

d)

FIG. 3. Effect of the current amplifier filter settings on the recorded current transient of a SY-PPV OLED. A constant forward bias of 3.73 Volts was applied to the device. (a) Plots of the relative current changes $\delta J_E / J_E$ follow a short microwave pulse of duration 400 ns and frequency 9.625 GHz, as a function of time after the pulse for three different band-pass settings. (b) The magnitudes of the FFT of the measured time-dependent currents for the three filter settings in (a). (c) Measurements of the current amplifier transfer functions for the three different filter settings as obtained by application of a well-defined oscillatory test signal with current $I = 18$ μA. (d) The plots of the data in (b) corrected by the transfer functions displayed in (c) show agreement within the given noise levels.

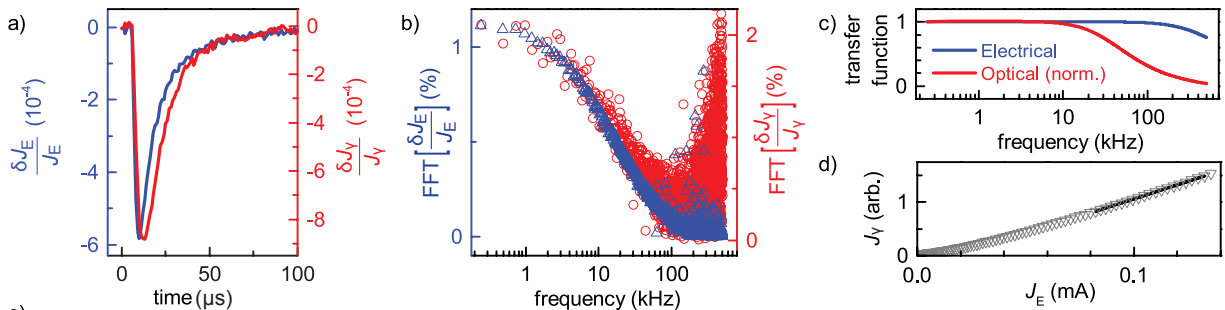

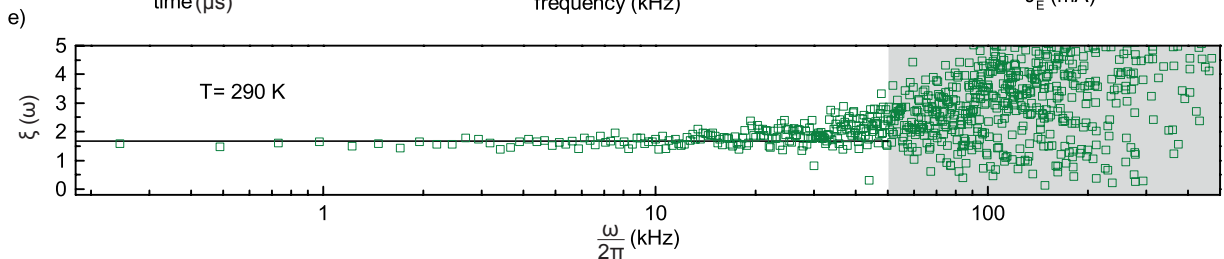

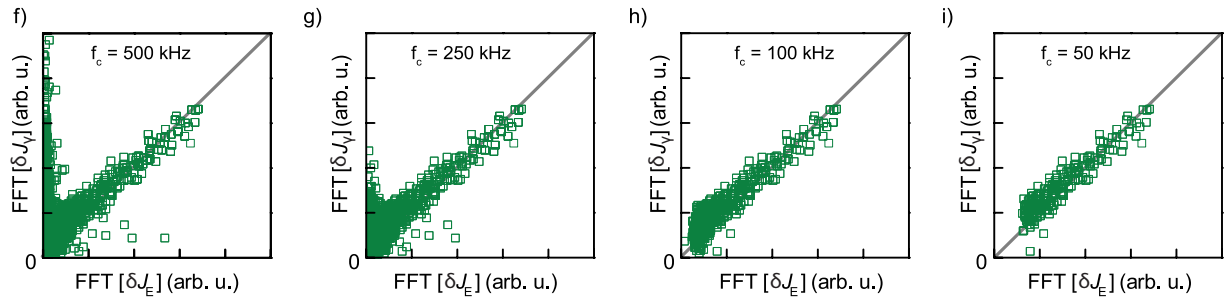

FIG. 4. Simultaneous room-temperature measurements pEDMR and pODMR of a SY-PPV OLED operating with a constant forward bias of 4 V. (a) Plots of the relative current changes $\delta J_{\rm E}/J_{\rm E}$ as well as the relative change of the detected EL intensity $\delta J_{\gamma}/J_{\gamma}$ following a short microwave pulse of duration 400 ns (frequency 9.6232 GHz), as a function of time after the pulse. (b) Plot of the magnitude of the FFT of the data in (a), corrected by the measured detection transfer functions shown in (c). (d) Plot of the detected light intensity as a function of current for the OLED studied. The black line is a linear fit to the data around the operating point which is used to obtain the dynamic detection efficiency $\alpha = \partial J_{\gamma}/\partial J_{\rm E}$. (e) Plot of the ratio $\xi(\omega) = \left|FFT\{\delta J_{\gamma}\}(\omega)\right|/\left|FT\{\delta J_{\rm E}\}(\omega)\right|/\left(\partial J_{\gamma}/\partial J_{\rm E}\right)$ as a function of frequency. (f) to (i) Correlation plots of the Fast Fourier components $\left|FFT\{\delta J_{\rm E}\}(\omega)\right|$ of the pEDMR measurements and the Fast Fourier components $\left|FFT\{\delta J_{\gamma}\}(\omega)\right|$ of the pODMR measurements shown in (b). Panel (f) displays all of the frequencies $\omega$ plotted in (e). In (g) to (i), data points at frequencies above the upper cut-off frequency $f_{\rm c}$, which decreases from 250 kHz in (g) to 50 kHz in (i), are excluded.

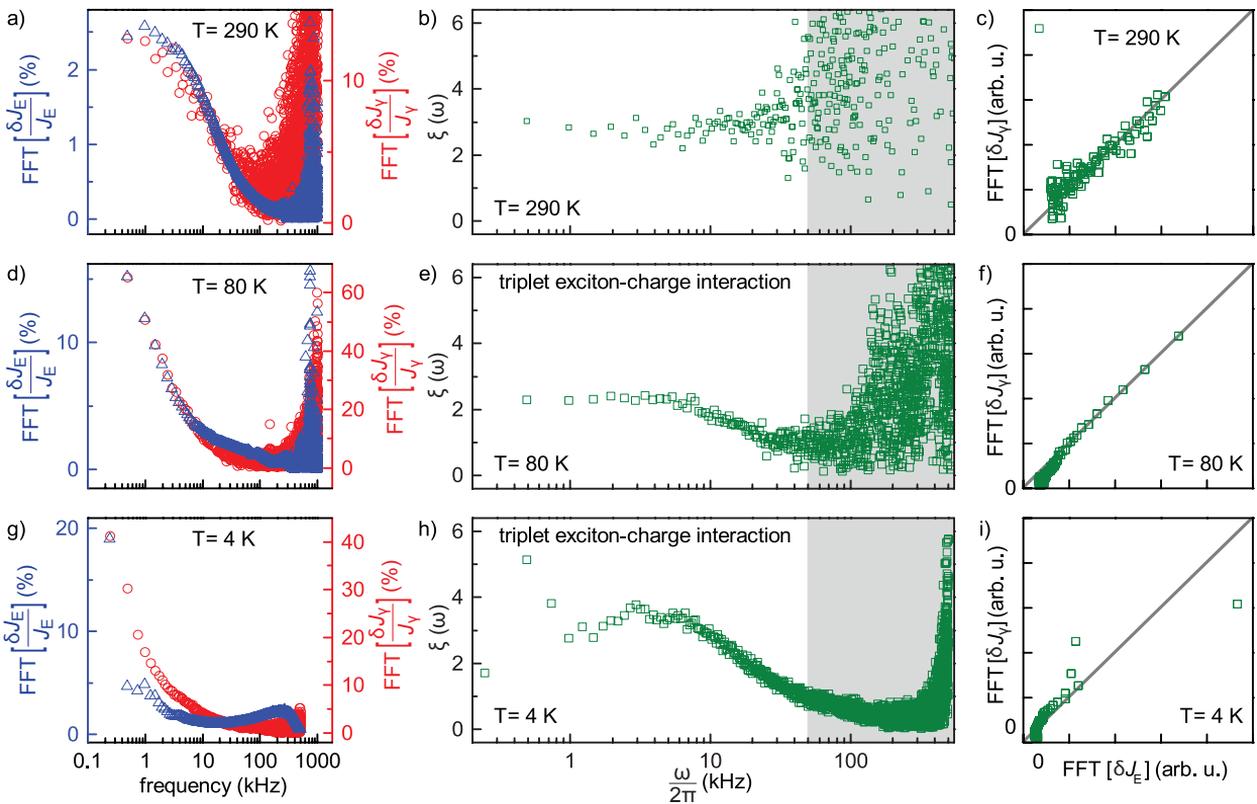

FIG. 5. Simultaneous room-temperature measurements of pEDMR and pODMR on a SY-PPV OLED at various temperatures. (a) Plot of the magnitude of the FFT of the transient relative current change $\delta J_E / J_E$ as well as the transient relative change of EL $\delta J_\gamma / J_\gamma$ following a short microwave pulse of duration 400 ns (frequency 9.6240 GHz). A constant forward bias is set to V = 13.7 V at $T$ = 100 K. (b) Plot of the ratio $\xi(\omega) = \left| FFT\{\delta J_\gamma\}(\omega) \right| / \left| FFT\{\delta J_E\}(\omega) \right| / \left( \partial J_\gamma / \partial J_E \right)$ as a function of frequency. (c) Correlation plot of the Fast Fourier components $\left| FFT\{\delta J_E\}(\omega) \right|$ of the pEDMR measurements with the Fast Fourier components $\left| FFT\{\delta J_\gamma\}(\omega) \right|$ of the pODMR measurements for each of the frequencies $\omega$ plotted in (b), except those in the grey shaded region. (d), (e), (f) Plots analogous to those in (a), (b), and (c) respectively, representing experiments conducted at 20 K with a bias voltage of 16.31 V. (g), (h), (i) Plots analogous to those in (a), (b), and (c), respectively, representing experiments conducted at 5K with a bias voltage of 16.6 V.

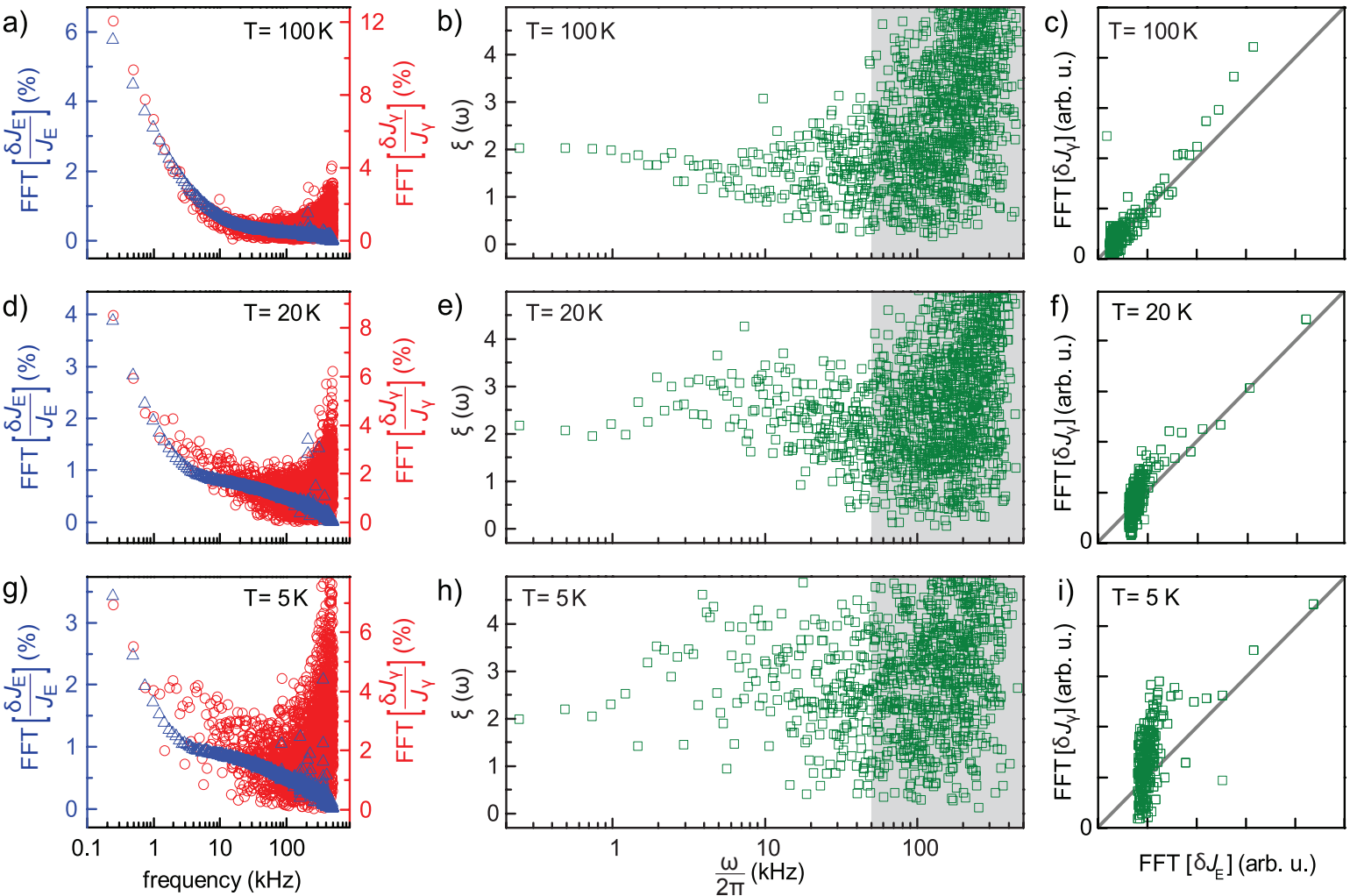

FIG. 6. Simultaneous room-temperature measurements of pEDMR and pODMR on an MEH-PPV OLED at various temperatures. (a) Plot of the magnitude of the FFT of the transient relative current change $\delta J_E / J_E$ as well as the transient relative change of EL $\delta J_\gamma / J_\gamma$ following a short microwave pulse of duration 400 ns (frequency 9.6232 GHz). The constant forward bias is set to V = 8 V at room temperature ($T$ = 290 K). (b) Plot of the ratio $\xi(\omega) = \left| FFT\{\delta J_\gamma\}(\omega) \right| / \left| FFT\{\delta J_E\}(\omega) \right| / \left( \partial J_\gamma / \partial J_E \right)$ as a function of frequency. (c) Correlation plot of the Fast Fourier components $\left| FFT\{\delta J_E\}(\omega) \right|$ of the pEDMR measurements and the Fast Fourier components $\left| FFT\{\delta J_\gamma\}(\omega) \right|$ of the pODMR measurements for each of the frequencies $\omega$ plotted in (b), excluding data points at frequencies in the grey shaded domain. (d), (e), (f) Plots analogous to those in (a), (b), (c) respectively, representing experiments conducted at 80 K with a bias voltage of 16.5 V. (g), (h), (i) Plots analogous to those in (a), (b), and (c), respectively, representing experiments conducted at 4 K with a bias voltage of 12.6 V. Whereas the ratio $\xi(\omega) = \left| FFT\{\delta J_\gamma\}(\omega) \right| / \left| FFT\{\delta J_E\}(\omega) \right| / \left( \partial J_\gamma / \partial J_E \right)$ is flat at room temperature, it clearly becomes frequency dependent at lower temperatures, indicating the involvement of an additional spin-dependent transport channel with a different frequency response. This additional channel is attributed to the interaction between triplet excitons and lone polarons.